\documentclass[]{interact}

\usepackage{epstopdf} 
\usepackage[caption=false]{subfig}
\usepackage[numbers,sort&compress]{natbib}

\usepackage{booktabs}
\usepackage{amsmath}
\usepackage{todonotes}

\usepackage{xcolor}

\bibpunct[, ]{[}{]}{,}{n}{,}{,}
\makeatletter
\def\NAT@def@citea{\def\@citea{\NAT@separator}}
\makeatother

\theoremstyle{plain}

\theoremstyle{definition}
\newtheorem{definition}{Definition}

\theoremstyle{remark}

\newtheorem{assumption}{Assumption}

\begin{document}


\title{Demonstration of a Time-Efficient Mobility System Using a Scaled Smart City}

\author{
\name{L.~E. Beaver, B. Chalaki, A.~M.~I. Mahbub, L. Zhao, R. Zayas, A.~A. Malikopoulos}
\affil{University of Delaware, Department of Mechanical Engineering, Newark, DE, USA 19716}
\thanks{CONTACT A.~A. Malikopoulos. Email: andreas@udel.edu}
}

\maketitle

\begin{abstract}
The implementation of connected and automated vehicle (CAV) technologies enables a novel computational framework to deliver real-time control actions that optimize travel time, energy, and safety. Hardware is an integral part of any practical implementation of CAVs, and as such, it should be incorporated in any validation method. However, high costs associated with full scale, field testing of CAVs have proven to be a significant barrier. In this paper, we present the implementation of a decentralized control framework, which was developed previously, in a scaled-city using robotic CAVs, and discuss the implications of CAVs on travel time. Supplemental information and videos can be found at \url{https://sites.google.com/view/ud-ids-lab/tfms}.
\end{abstract}


\section{Introduction}
\begin{figure}
    \centering
    \includegraphics[width=.8\linewidth]{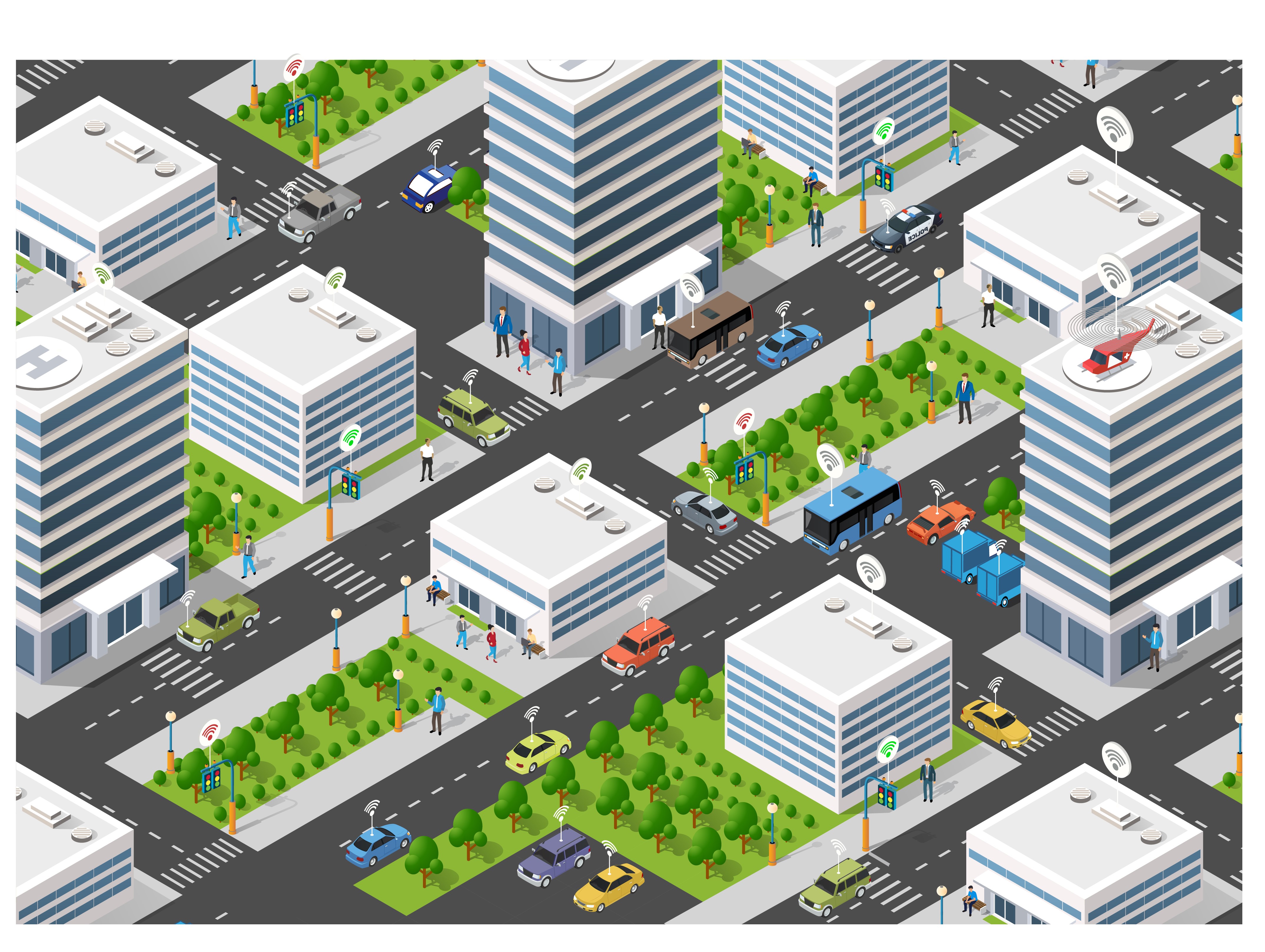}
    \caption{A city enabled by connectivity and automation technologies.}
    \label{fig:smart_city}
\end{figure}

Connectivity and automation provide the most intriguing opportunity for enabling users to better monitor transportation network conditions and make better operating decisions to reduce energy consumption, greenhouse gas emissions, travel delays, and improve safety. In the context of a smart city, wireless connectivity provides free-flow of information among entities, while automation provides precise execution upon such available information for moving goods and people safely and efficiently (Fig. \ref{fig:smart_city}). The availability of vehicle-to-vehicle (V2V) and vehicle-to-infrastructure (V2I) communication has the potential to ease congestion and improve safety by enabling vehicles to respond rapidly to changes in their mutual environment. Furthermore, vehicle automation technologies can aim at developing robust vehicle control systems that can quickly respond to dynamic traffic operating conditions.

With the advent of emerging information and communication technologies, we are witnessing a massive increase in the integration of our energy, transportation, and cyber networks. 
These advances, coupled with human factors, are giving rise to a new level of complexity in transportation networks  \cite{Malikopoulos2016c}. As we move to increasingly complex emerging transportation systems, with changing landscapes enabled by connectivity and automation, future transportation networks could shift dramatically with the large-scale deployment of connected and automated vehicles (CAVs). On the one hand, with the generation of massive amounts of data from vehicles and infrastructure, there are opportunities to develop optimization methods to identify and realize a substantial energy reduction of the transportation network, and to optimize the large-scale system behavior using the interplay among vehicles. On the other hand, evaluation and validation of new control approaches under different traffic scenarios is a necessity to ensure successful implementation per vehicle alongside desired system-level outcomes. 

\begin{figure}
    \centering
    \includegraphics[width=.8\linewidth]{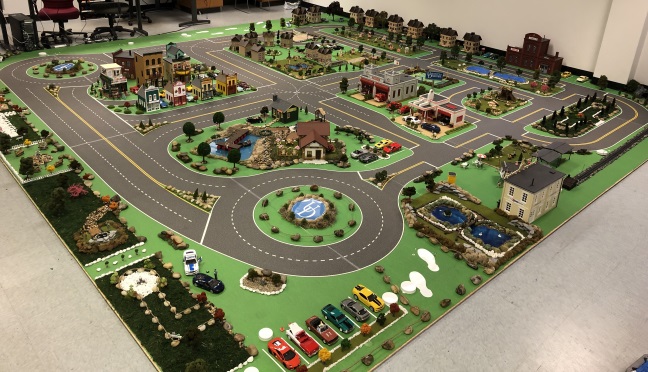}
    \caption{The University of Delaware's Scaled Smart City.}
    \label{fig:udssc}
\end{figure}

The overarching goal of this paper is the experimental demonstration of a decentralized control framework for CAVs presented in \cite{Zhao2018} using the University of Delaware's Scaled Smart City (UDSSC). UDSSC is a $1$:$25$ scaled testbed representing an urban environment with robotic CAVs that can replicate real-world traffic scenarios in a controlled environment (Fig. \ref{fig:udssc}). UDSSC can be used to explore the acquisition and processing of V2V and V2I communication. It can also be used to validate control algorithms for CAV coordination in specific transportation segments, e.g., intersections, merging roadways, and roundabouts, by mitigating the high costs and safety concerns associated with real-world field testing of CAVs. As an intermediate scale testbed, the UDSSC is an ideal platform to gain insight into the execution of high-level planning and coordination on physical hardware with noise, disturbances, and communication delays. In this paper, our emphasis is on the generation of energy-optimal trajectories, and as such, we do not consider the problem of describing low-level controllers which track the optimal trajectories.

The structure of the paper is organized as follows. In Section \ref{sec: literature}, we discuss related work on the optimal control for CAVs reported in the literature. In Section \ref{sec: framework}, we review the decentralized control framework presented in \cite{Zhao2018} for the  coordination of CAVs in a transportation network. Then, we describe briefly the UDSSC testbed in Section \ref{sec:setup}, and present simulation and experimental results in Section \ref{sec: results}. Finally, we draw concluding remarks from the experiments in Section \ref{sec: conclusion}.

\section{Related Work} \label{sec: literature}
CAVs have attracted considerable attention for the potential of improving mobility and safety along with energy and emission reduction \cite{Spieser2014, Fagnant2014}. There have been two major approaches to utilizing connectivity and automation to improve transportation efficiency and safety, namely, platooning and traffic smoothing. 

The first approach utilizes connectivity and automation to form closely-coupled vehicular platoons to  reduce aerodynamic drag effectively, especially at high cruising speeds. The concept of forming platoons of vehicles 
was a popular system-level approach to address traffic congestion, 
which gained momentum in the 1980s and 1990s \cite{Shladover1991,Rajamani2000}. Such automated transportation system can alleviate congestion, reduce energy use and emissions, and improve safety while increasing throughput significantly. The Japan ITS Energy Project \cite{tsugawa2013}, the Safe Road Trains for the Environment program \cite{davila2010sartre}, and the California Partner for Advanced Transportation Technology \cite{shladover2007path} are among the mostly-reported efforts in this area.

The second approach is to smooth the traffic flow by \emph{centralized} or \emph{decentralized} vehicle control to reduce spatial and temporal speed variation and braking events, e.g., automated intersection crossing \cite{Lee2012, rakha2011eco, Malikopoulos2017, mahbub2019ACC, chalaki2019a}, cooperative merging \cite{Rios-Torres2017,Ntousakis2016aa, Zhao2018}, and speed harmonization through optimal vehicle control \cite{Malikopoulos2018c}. In centralized approaches, there is at least one task in the system that is globally decided for all vehicles by a single central controller, whereas in decentralized approaches, the vehicles are treated as autonomous agents that collect traffic information to optimize their specific performance criteria while satisfying physical constraints. One of the very early efforts in this direction was proposed by Athans \cite{Athans1969} for safe and efficient coordination of merging maneuvers with the intention to avoid congestion. Since then, numerous approaches have been proposed on coordinating CAVs to improve traffic flow \cite{Kachroo1997, Antoniotti1997, Ran1999}, and to achieve safe and efficient control of traffic through various traffic bottlenecks where potential vehicle collisions may happen \cite{Dresner2004,Dresner2008,DeLaFortelle2010, Huang2012,Zohdy2012,Yan2009,Li2006,Zhu2015a,Wu2014,kim2014}. In terms of energy impact, many studies have shown that significant fuel consumption savings could be achieved through eco-driving and vehicle optimal control without sacrificing driver safety \cite{Barth2009, Berry2010, wu2011, Malikopoulos2017, Ntousakis2016aa, Zhao2018}. Considering near-future CAV deployment, recent research work has also explored both  traffic and energy implications of partial penetration of CAVs under different transportation scenarios, e.g., \cite{Malikopoulos2018a, Rios2018, zhong2017}. Several survey papers that report the research efforts in this area 
can be found in \cite{Malikopoulos2016a, guanetti2018, wang2018r}.

Although previous work has shown promising results emphasizing the potential benefits of coordination of CAVs, validation has been primarily in simulation.
Some progress has been made with constructing experimental testbeds, such as MIT's Duckietown \cite{Paull2017}, which focuses primarily on local perception and autonomy, and the Cambridge Minicars \cite{Hyldmar2019}, which is a testbed for cooperative driving in highway conditions. In contrast, the UDSSC focuses on traffic coordination in an urban system, where a majority of stop-and-go driving occurs.
In previous work, we presented the experimental validation of the solution to the unconstrained merging roadway problem in UDSSC using $10$ robotic CAVs \cite{Stager2018}. In this paper, we demonstrate the impact of an optimal decentralized framework, developed in earlier work \cite{Zhao2018}, for coordinating CAVs in a transportation network with multiple conflict zones where a 
lateral collision may occur.

\section{Decentralized Control Framework} \label{sec: framework}
\begin{figure}[ht]
    \centering
    \includegraphics[width=.8\linewidth]{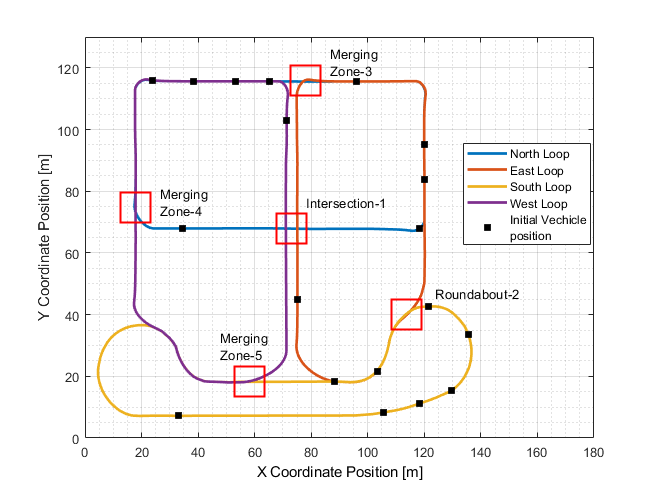}
    \caption{Vehicle routes in the University of Delaware's Scaled Smart City environment.}
    \label{fig:udssc_vissim}
\end{figure}

We consider a network of CAVs driving in the roadway network in UDSSC, which consists of several \textit{conflict zones}, e.g., ramps, roundabouts, and intersections, where lateral collisions may occur (marked with red boxes in Fig. \ref{fig:udssc_vissim}). For each conflict zone, there is a coordinator that communicates with all CAVs traveling within 
its communication range. In practice, the coordinator can be stationary roadside units for general-purpose traffic monitoring and message dissemination or mobile roadside units (e.g., transit vehicles, drones), which can provide dynamic traffic monitoring and communication support at specific locations or along corridors. Each CAV is retrofitted with a communication device necessary to interact with other vehicles and local infrastructure within their communication range. Upstream of a conflict zone, we define a \textit{control zone} in each direction, inside of which, the CAVs coordinate with each other in order to travel through the conflict zone without any collisions (Fig. \ref{fig:udssc_vissim}). The length of the control zone can be considered to be the maximum communication range of a V2X device. Outside the control zone, the CAVs behave as human-driven vehicles. For simplicity, we do not consider multi-lane trajectories or any lane changes within the control zone. This is the focus of ongoing work \cite{Malikopoulos2019b}, and it is not discussed here.

\subsection{Modeling Framework and Assumptions}
Let $z \in \mathcal{Z}$ be the index of a conflict zone in the corridor. Let $\mathcal{N}_z(t) = \{1,2,..., N(t)\}$ be a queue of CAVs to be analyzed corresponding to the conflict zone $z$, where $N(t)\in\mathbb{N}$ is the total number of CAVs at the time $t\in\mathbb{R}^{+}$. The dynamics of each vehicle $i\in \mathcal{N}_z(t)$, are represented with a state equation
\begin{equation} \label{eq:state}
\dot{\mathbf{x}}(t) = f(t, \mathbf{x}_i, u_i), ~ \mathbf{x}_i(t_i^{z,0}) = x_i^{z,0},
\end{equation}
where $\mathbf{x}_i(t), u_i(t)$ are the state of the vehicle and control input, $t_i^{z,0}$ is the initial time of vehicle $i\in \mathcal{N}_z(t)$ entering the control zone corresponding to the conflict zone $z$, and $\mathbf{x}_i^{z,0}$ is the value of the initial state. For simplicity, we model each vehicle as a double integrator, i.e., $\dot{p}_i = v_i(t)$ and $\dot{v}_i = u_i(t)$, where $p_i(t) \in \mathcal{P}_i, v_i(t) \in \mathcal{V}_i$, and $u_i(t) \in \mathcal{U}_i$ denote the position, speed, and acceleration/deceleration (control input) of each vehicle $i$. Let $\mathbf{x}_i(t)=[p_i(t)~ v_i(t)]^T$ denote the state of each vehicle $i\in \mathcal{N}_z(t)$, with initial value $\mathbf{x}_i^{z,0}(t)=[0 ~ v_i^{z,0}(t)]^T$, taking values in the state space $\mathcal{X}_i=\mathcal{P}_i\times  \mathcal{V}_i$. The sets $\mathcal{P}_i, \mathcal{V}_i$, and $\mathcal{U}_i, i\in \mathcal{N}(t)$, are complete and totally bounded subsets of $\mathbb{R}$. The state space $\mathcal{X}_i$ for each vehicle $i$ is closed with respect to the induced topology on $\mathcal{P}_i\times  \mathcal{V}_i$ and thus, it is compact.

To ensure that the control input and vehicle speed are within a given admissible range, we impose the constraints 
\begin{align}\label{eq:constraints}
& u_{min} \leq u_i(t) \leq u_{max},~ \text{and} \\
& 0 \leq v_{min} \leq v_i(t) \leq v_{max}, ~ \forall t \in [t_i^{z,0}, ~ t_i^{z,f}], \nonumber
\end{align}
where $u_{min}, u_{max}$ are the minimum deceleration and maximum acceleration respectively, $v_{min}, v_{max}$ are the minimum and maximum speed limits respectively, and $t_i^{z,0}$, $t_i^{z,f}$ are the times that each vehicle $i$ enters and exits the conflict zone $z$.

To 
avoid a rear-end collision between two consecutive vehicles traveling in the same lane, the position of the preceding vehicle should be greater than, or equal to the position of the following vehicle plus a predefined safe distance $\delta_i(t)$, 
which is proportional to the speed of vehicle $i$, $v_i(t)$. Thus, we impose the rear-end safety constraint
\begin{equation} \label{eq:safety}
s_i(t)=p_k(t)-p_i(t)\geq \delta_i(t),\quad\forall t \in  [t_i^{z,0}, ~ t_i^{z,f}],
\end{equation}
where vehicle $k$ is immediately ahead of $i$ on the same lane. 
The minimum safe distance, $\delta_i(t)$, is a function of speed, namely
\begin{gather}
    \delta_i(t) = \gamma_i + h\cdot v_i(t),
\end{gather}
where $\gamma_i$ is the standstill distance and $h$ is the minimum safe time gap that CAV $i$ can maintain while following another vehicle. 

\begin{definition}
For each CAV $i\in\mathcal{N}_z(t)$
approaching a conflict zone $z\in\mathcal{Z}$, we define two
subsets of $\mathcal{N}_z(t)$ depending on 
the physical location 
of $i$ 
inside the control zone:
1) $\mathcal{L}_{i}^{z}(t)$ contains all CAVs traveling in  the same road and lane as CAV $i$, which may cause rear-end collision with CAV $i$, 
and 2) $\mathcal{C}_{i}^{z}(t)$ contains all CAVs traveling on a different road from $i$ that can 
cause a lateral collision inside the conflict zone $z$. \label{def:1}
\end{definition}

\begin{definition}
For each vehicle $i\in\mathcal{N}_z(t)$, we define the set $\Gamma_{i}^{z}$ that consists of 
the positions along the lane where a lateral collision is possible in a conflict zone $z$, namely
\begin{equation}
\Gamma_{i}^{z}\triangleq\Big\{p_i(t)~|~t\in\lbrack t_{i}^{z,m},t_{i}^{z,f}]\Big\},
\label{eq:def1}
\end{equation}
where  $t_{i}^{z,m}$ is the time that vehicle $i$ exits the control zone (and enters the conflict zone $z$), and $t_{i}^{z,f}$ is the time that vehicle $i$ exits the conflict zone $z$.
\end{definition}

Consequently, to avoid a lateral collision for any two vehicles $i,j\in \mathcal{N}_z(t)$ on different roads we impose 
the following constraint
\begin{equation}
\Gamma_{i}^{z}\cap\Gamma_{j}^{z}=\varnothing, \text{ \ }j\in\mathcal{C}_{i}^z(t) \text{, \ \ \ }\forall t\in\lbrack
t_{i}^{z,m},t_{i}^{z,f}]. \label{eq:lateral}%
\end{equation}

The above constraint implies that only one vehicle at a time can be inside the conflict zone $z$ when there is a potential for a lateral collision. If the length of the conflict zone is long, then this constraint 
may dissipate the 
capacity of the road. However, the constraint is not restrictive in the problem formulation, and it can be modified appropriately.

In the modeling framework described above, we impose the following assumptions:

\begin{assumption}
For each CAV $i$, none of the constraints is active at $t_i^{z,0}$ for conflict zone $z$.
\end{assumption}

\begin{assumption}
Each CAV $i$ 
can communicate with the coordinator and other CAVs to receive local information without errors or delays. 
\end{assumption}

The first assumption ensures that the initial state and control input are feasible. The second assumption might be strong, but it is relatively straightforward to relax as long as the noise in the measurements and/or delays is bounded. For example, we can determine upper bounds on the state uncertainties as a result of sensing or communication errors and delays, and incorporate these into more conservative safety constraints.

\subsection{Communication Structure}
When a CAV $i\in\mathcal{N}_z(t)$ enters a control zone, it communicates with a coordinator, assigned to the corresponding conflict zone, and the other CAVs inside the control zone. Note that the coordinator is not involved in any decision for the CAVs, and it only facilitates the communication of appropriate information among vehicles. The coordinator handles the information between the vehicles as follows. When a vehicle enters the control zone of the conflict zone $z$  at time $t$, the coordinator assigns a \textit{unique identity} $i=N(t)+1$, which is an integer representing the location of the vehicle in a first-in-first-out queue $\mathcal{N}_z(t)$. If any two, or more, vehicles enter the control zone at the same time, then the coordinator randomly selects their positions in the queue. 

\begin{definition}
For each CAV $i\in\mathcal{N}_z(t)$ entering the control zone $z$, the \textit{information set} $Y_{i}^{z}(t),$ is defined as
\begin{gather}
Y_{i}^{z}(t) \triangleq\Big\{p_{i}(t),v_{i}(t),\mathcal{L}_i^{z}(t),\mathcal{C}_i^{z}(t),t_{i}^{z,m}\Big\},\forall t \in\lbrack t_{i}^{z,0},t_{i}^{z,m}],
\end{gather}
\label{infoset}where $p_{i}(t),v_{i}(t)$ are the position and speed of CAV $i$ inside the control zone $z$, $t_{i}^{z,0}$ is the time when vehicle $i$ enters the control zone for conflict zone $z$, and $t_{i}^{z,m}$ is the time targeted for vehicle $i$ to enter the conflict zone $z$. The set $Y_{i}^{z}(t)$ includes all information that each vehicle shares.
\end{definition}

The time $t_{i}^{z,m}$ that the vehicle $i$ will be entering the conflict zone $z$ maximizes the throughput while considering the maximum and minimum speed limits. 
Therefore, to ensure that \eqref{eq:safety} and \eqref{eq:lateral} are satisfied at $t_{i}^{z,m},$ we impose the following conditions which depend on the subset that the vehicle $i-1\in\mathcal{N}_z(t)$ belongs to.

If CAV $i-1\in\mathcal{L}_{i}^{z}(t)$,
\begin{equation}
t_{i}^{z,m}=\max\Bigg\{ \min\Big\{t_{i-1}^{z,m}+\frac{\delta(t)}{v^{z}},\frac{L^z}{v_{min}}\Big\}, \frac{L^z}{v_{i}(t_{i}^{z,0})}, \frac{L^z}{v_{max}}\Bigg\}.\label{eq:condition1a}%
\end{equation}

If CAV $i-1\in\mathcal{C}_{i}^{z}(t)$,
\begin{equation}
t_{i}^{z,m}=\max\Bigg\{ \min\Big\{t_{i-1}^{z,m}+\frac{S^z}{v^{z}},\frac{L^z}{v_{min}}\Big\}, \frac{L^z}{v_{i}(t_{i}^{z,0})}, \frac{L^z}{v_{max}}\Bigg\},\label{eq:condition1b}%
\end{equation}
where $S^z$ is the length of conflict zone $z$, $L^{z}$ is the length of control zone for zone $z$, $v^{z}$ is the constant imposed speed inside the conflict zone $z$, and $v_{i}(t_{i}^{z,0})$ is the initial speed of vehicle $i$ when it enters the control zone at $t_{i}^{z,0}$. The conditions \eqref{eq:condition1a} and \eqref{eq:condition1b} ensure that the time $t_{i}^{z,m}$ each vehicle $i$ will be entering the conflict zone is feasible and can be attained based on the imposed speed limits inside the control zone. In addition, for low traffic flow where vehicles $i-1$ and $i$ might be located far away from each other, there is no compelling reason for vehicle $i$ to accelerate within the control zone to have a distance $\delta(t)$ from vehicle $i-1$, if $i-1\in\mathcal{L}_{i}^{z}(t)$, or a distance $S^z$ if $i-1\in\mathcal{L}_{i}^{z}(t)$, at the time $t_{i}^{z,m}$ that vehicle $i$ enters the conflict zone $z$. Therefore, in such cases, vehicle $i$ can keep cruising within the control zone with the initial speed $v_{i}(t_{i}^{z,0})$ that entered the control zone at $t_{i}^{z,0}.$ 

The recursion is initialized when the first vehicle enters
the control zone $z$, i.e., it is assigned $i=1$. In this case, $t_{1}^{z,m}$ can be externally assigned as the desired exit time of this vehicle whose behavior is unconstrained. Thus, the time $t_{1}^{z,m}$ is fixed and available through $Y_{1}(t)$. The second vehicle will access $Y_{1}^{z}(t)$ from vehicle 1 to compute the time $t_{2}^{z,m}$. The third vehicle will access $Y_{2}^{z}(t)$ from vehicle 2, and the communication process will continue with the same fashion until vehicle $N(t)$ in the queue accesses $Y_{N(t) -1}^{z}(t)$.

\subsection{Optimal Control Problem Formulation}

By controlling the 
entry time of the vehicles, the speed of queue build-up at each conflict zone decreases. Thus, the congestion recovery time is also reduced -- the latter results in maximizing the throughput in the conflict zone.
We now consider the problem of deriving the optimal control input (acceleration/deceleration) of each CAV inside each control zone separately under hard safety constraints to avoid collisions. Moreover, by optimizing the acceleration/deceleration of each vehicle, we minimize transient operation. This will 
have direct benefits in energy consumption since the vehicles are optimized to travel over steady state operating points (constant torque and speed) \cite{Malikopoulos2008b}.

Since the coordinator for a conflict zone $z$ is not involved in any decision on the vehicle coordination, we formulate the following optimization problem for each vehicle in the queue upstream of conflict zone $z$, the solution of which can be implemented in real-time 
\begin{gather} \label{eq:decentral}
\min_{u_i}\frac{1}{2}\int_{t_i^{z, 0}}^{t_i^{z,m}} u_i^2(t)dt, ~z \in \mathcal{Z}, \\
\text{Subject to}: (\ref{eq:state}), (\ref{eq:constraints}),
p_{i}(t_i^{z,0})=p_{i}^{z,0}\text{, }v_{i}(t_i^{z,0})=v_{i}^{z,0}\text{, }p_{i}(t_i^{z,m})=p_{z},\nonumber\\
\text{and given }t_i^{z,0}\text{, }t_i^{z,m},\nonumber
\end{gather}
where $p_{z}$ is the location (i.e., entry position) of the conflict zone $z$, $t_i^{z,m}$ is the time that the vehicle $i$ enters the conflict zone, and $p_{i}^{z,0}$, $v_{i}^{z,0}$ are the initial position and speed of vehicle $i\in\mathcal{N}_z(t)$ when it enters the control zone of conflict zone $z$. By minimizing the $L^2$ norm of acceleration we minimize transient engine operation which results in an overall improvement in energy efficiency.

For the analytical solution and real-time implementation of the control problem \eqref{eq:decentral}, we apply Hamiltonian analysis. The analytical solution of \eqref{eq:decentral} without considering state and control constraints was presented in earlier work
 \cite{Rios-Torres2017,Ntousakis2016aa} for coordinating CAVs in real-time at highway on-ramps. When the state and control constraints are not active, the optimal control input (acceleration/deceleration) as a function
of time is given by 
\begin{equation}
u_{i}^{*}(t)=a_{i}t+b_{i}, ~t_{i}^{z,0}\le t\le t_{i}^{z,m} \label{eq:20},
\end{equation}
and the optimal speed and position for
each vehicle are 
\begin{equation}
v_{i}^{*}(t)=\frac{1}{2}a_{i}t^{2}+b_{i}t+c_{i}, ~ t_{i}^{z,0}\le t\le t_{i}^{z,m}, \label{eq:21}
\end{equation}
\begin{equation}
p_{i}^{*}(t)=\frac{1}{6}a_{i}t^{3}+\frac{1}{2}b_{i}t^{2}+c_{i}t+d_{i}, ~ t_{i}^{z,0}\le t\le t_{i}^{z,m}, \label{eq:22}
\end{equation}
where $a_{i}$, $b_{i}$, $c_{i}$ and $d_{i}$ are constants of integration that
can be computed by using the initial and final conditions. 
Similar results to \eqref{eq:20}-\eqref{eq:22} can be obtained when
the state and control constraints become active within the control zone. In this case, the constrained and unconstrained arcs need to be pieced together to satisfy the Euler-Lagrange equations and the necessary conditions of optimality. The different cases of the state and control constraint activation along with the corresponding solution can be found in \cite{Malikopoulos2017}, whereas the complete analytical solution that includes the rear-end safety constraint is reported in \cite{Malikopoulos2019a}. In the present work, we do not consider any constrained optimization cases as none of the constraints in \eqref{eq:constraints} become active within the optimal control path during the simulation, as shown in Section \ref{sec: results}.

 \section{Simulation and Experimental Environment}\label{sec:setup}

\subsection{Simulation Setup}\label{sec:simulation-setup}
To implement the control framework presented in the previous section, and to generate the input information required for UDSSC, we first use the microscopic multi-modal simulation platform PTV VISSIM. We create a simulation setup replicating the UDSSC map and define a network consisting of four different looped routes and five bottlenecks (one intersection, one roundabout, and three merging scenarios), as shown in Fig. \ref{fig:udssc_vissim}. In order to maintain compatibility with the UDSSC experiment, we design each of the routes to hold a finite number of vehicles ($19$ vehicles in total) traveling in loops for finite simulation run-time. Among the $19$ vehicles, we consider $9$ vehicles as the target (ego-vehicles) to evaluate their performance metrics in different scenarios. We use the rest of the vehicles to increase the traffic volume in the urban network and create congestion in the baseline scenario. The vehicles maintain a low desired speed of 7 m/s for their uncontrolled urban commute throughout the network. Therefore, the desired speed at all exits of the control zones is set to be equal to the urban speed. We select the maximum and minimum allowable speed of  8.33 m/s and 2 m/s, respectively. The maximum and minimum acceleration of the vehicles was taken as 3 m/s$^2$ and -3 m/s$^2$, respectively. To evaluate the effectiveness of the proposed optimal vehicle dynamics control, we consider two different cases:

\textit{a. Baseline Scenario:} We construct the baseline scenario by considering all vehicles as human-driven and without any V2V communication capability. The vehicles subscribe to the Wiedemann car following model. The Wiedemann car following model is a psycho-physical model to emulate the driving behavior of real human-driven vehicles. The model was first presented in 1974 by Wiedemann \cite{Wiedemann1974} and has been adopted by the software PTV-VISSIM as one of its in-built car following models. The complete detail of this elaborate model can be found in the literature \cite{Wiedemann1974}. We do not change the VISSIM default parameters of the Wiedemann model to study their impact on vehicle behavior and traffic flow, as 
such exposition falls outside the scope of this paper. 
We build a fixed time signalized intersection for the four-way traffic at the center. The fixed time signalized intersection in the simulation provides Signal Phase and Timing (SPaT) messages to the leading vehicle internally through VISSIM. Based on the SPaT information, the leading vehicle stops at the desired position by adopting the Wiedemann car following model. We adopt priority-based (yield/stop) movement for the other four waypoints consisting of the roundabout and merging scenarios, where the secondary-road vehicles yield to the main-road vehicles. In both cases, VISSIM uses the “approaching point” parameter of Wiedemann car following model to detect any approaching obstacles and apply necessary braking to slow down or stop.

\textit{b. Optimal Controlled Scenario:} In the optimal scenario, all 
19 vehicles follow our optimal control framework. The vehicles are connected with each other inside the control zone through V2V communication capability and are automated within the control zone. Therefore, they can plan their optimal path inside the control zone, avoiding any lateral or rear-end collisions while optimizing their own travel time and fuel efficiency. In this scenario, we do not consider the fixed-time signal and movement priorities considered in the baseline case. We consider five isolated coordinators with a control zone of  45 m for each conflict zone (see Fig. \ref{fig:udssc_vissim}). For the uncontrolled paths in-between the control zones, the vehicles adopt the Wiedemann car following model \cite{Wiedemann1974} to traverse their respective routes.

\subsection{University of Delaware's Scaled Smart City} \label{sec: udssc}
UDSSC is a $1$:$25$ scaled testbed spanning over $400$ square feet (see Fig. \ref{fig:udssc}) and is capable of accommodating scaled robotic CAVs. It is equipped with a VICON motion capture system that uses eight cameras to track the position of each vehicle with sub-millimeter accuracy. Each road in the UDSSC is built up from arc or line segments. In order to track the desired vector position of each CAV, all road segments are parameterized in terms of their total length. This formulation allows each vehicle to calculate its desired position in UDSSC based only on the scalar distance along its current path, which is achieved by numerically integrating the speed profile in real-time on both the mainframe computer and each CAV. This decoupling of speed and position allows significant flexibility in UDSSC, especially in dynamic-routing scenarios.

\subsubsection{Connected and Automated Vehicles}
\begin{figure}
    \begin{minipage}{\linewidth}
    \includegraphics[height=2.5in]{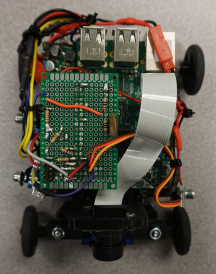}\hfill
    \includegraphics[height=2.5in]{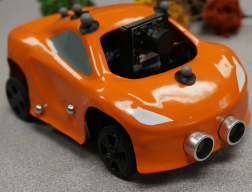}%
    \end{minipage}
    \caption{A picture of the connected and automated vehicle's electronics (left) and outer shell with VICON markers, ultrasonic sensors, and camera visible (right).}
    \label{fig:CAV}
\end{figure}

The CAVs of UDSSC  (see Fig. \ref{fig:CAV}) have been designed using off the shelf electrical components and $3$D printed parts created at the University of Delaware. The primary microcontroller on the CAV is a Raspberry Pi $3$B running Ubuntu Mate and ROS Kinetic. An Arduino Nano is used as a slave processor for the Pi to do low-level motor control and ad-hoc analog to digital conversion for the state of charge (SOC) measurements. The CAV's rear-wheel drive train is powered by a Pololu $75.8$:$1$, $6$ V micro metal gearmotor; the motor is controlled using a motor controller, and encoder for feedback, through the Arduino. Power from the gearmotor is transferred to the rear axle with two $3$D printed gears with a $1$:$1$ ratio, and two rubberized wheels with radius $r=1.6$ cm are mounted directly to the rear axle. The motor controller receives power through a $5$ V regulator, and a pulse-width modulated command from the Arduino is used to control the motor's speed. Steering is achieved by a custom $3$D printed Ackermann-style steering mechanism actuated by a Miuzei micro servo motor, which again is controlled directly by the Arduino. The CAVs are also equipped with a Pi Camera, ultrasonic sensors, and a SOC measurement circuit to collect experimental data and reduce the overall reliance on VICON. A power regulator manages the voltage requirement of the Pi and Arduino by supplying a regulated $5$ V DC from two $3000$ mAh $3.7$ V Li-ion batteries configured in series. With this hardware configuration, the CAV is able to run and collect experimental data at $20$ Hz for up to $2$ hours.

\subsubsection{Control System Architecture}
Coordination of the CAVs within the UDSSC is achieved using a multi-level control framework spanning a central \textit{mainframe} computer (Processor: Intel Core i7-6950X CPU @ $3.00$ GHz x $20$, Memory: $125.8$ Gb) and the individual CAVs in the experiment  (Raspberry Pi 3B). The mainframe runs an Ubuntu 16.04.5 LTS Linux distribution and ROS Kinetic. High-level routing is achieved by a multithreaded C++ program running on the mainframe computer. For this set of experiments, the mainframe is initialized with the path information and speed profiles for each CAV. At the start of the experiment, each CAV sets its temporal baseline from which it measures all later times; this avoids the problem of synchronizing CAV clocks, as all information is calculated relative to the experiment start time. During the experiment, the mainframe passes a message to each CAV containing its current position and two seconds of trajectory data using the UDP/IP protocol at $50$ Hz. The CAV receives trajectory information from the mainframe and uses a modified Stanley \cite{Thrun2007Stanley:Challenge} controller to handle lane tracking, while a feedforward-feedback \cite{Spong2004RobotEdition} PID controller tracks the desired speed profile.

Medium and low-level control is accomplished onboard each CAV in a purely distributed manner. Using information from the mainframe, each 
CAV updates at $50$ Hz to calculate a lateral, heading, and distance error. The lateral and heading errors are then passed to the Stanley controller to calculate an output steering angle. Meanwhile, the position error and desired speed are used in a feedforward-feedback controller to calculate the desired motor speed. The desired speed and steering angle are then passed to the Arduino Nano, which runs a low-level PID controller to precisely control the gearmotor and steering servo.

\subsection{Experimental Setup}
In order to replicate the simulation results in the UDSSC, speed profiles for the 9 ego vehicles  are exported from VISSIM 
to the mainframe. The path information and speed profile are dispersed to each CAV for the duration of the experiment. Then, the CAVs at UDSSC 
numerically integrate the speed profile data in real-time to calculate their desired position, allowing them to track the desired speed and position in a decentralized manner. 
Simultaneously, the mainframe computer integrates the speed profile in order to send current and future path information to the CAVs.


\section{Results} \label{sec: results}

\subsection{Simulation Results}
\begin{figure}[htht] 
\centering
  \subfloat[North Loop\label{fig:5-rt1}]{
    \includegraphics[width=0.48\textwidth]{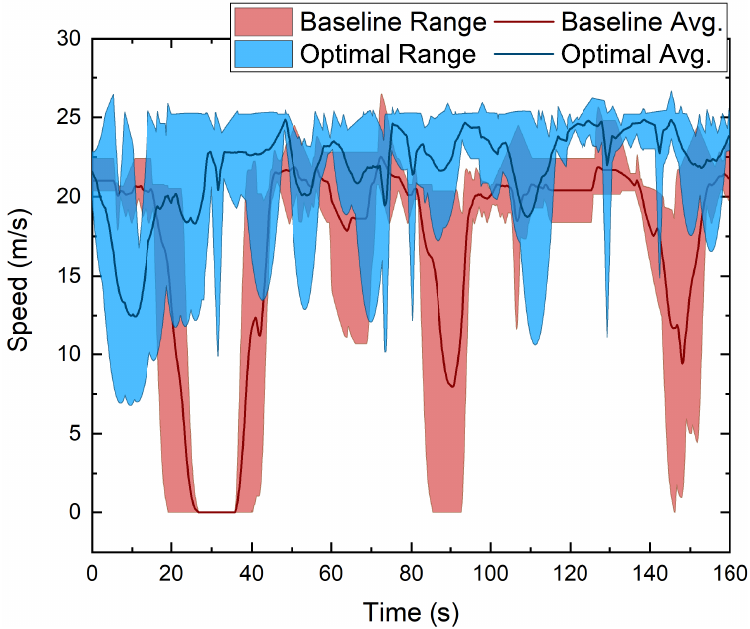}
  } 
  \subfloat[East Loop\label{fig:5-rt2}]{
    \includegraphics[width=0.48\textwidth]{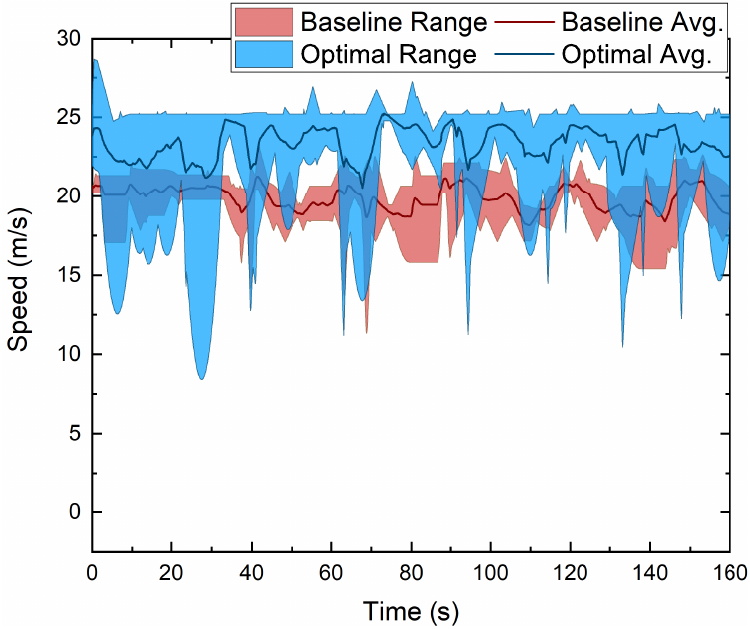}
  }\\
  \subfloat[South Loop\label{fig:5-rt3}]{
    \includegraphics[width=0.48\textwidth]{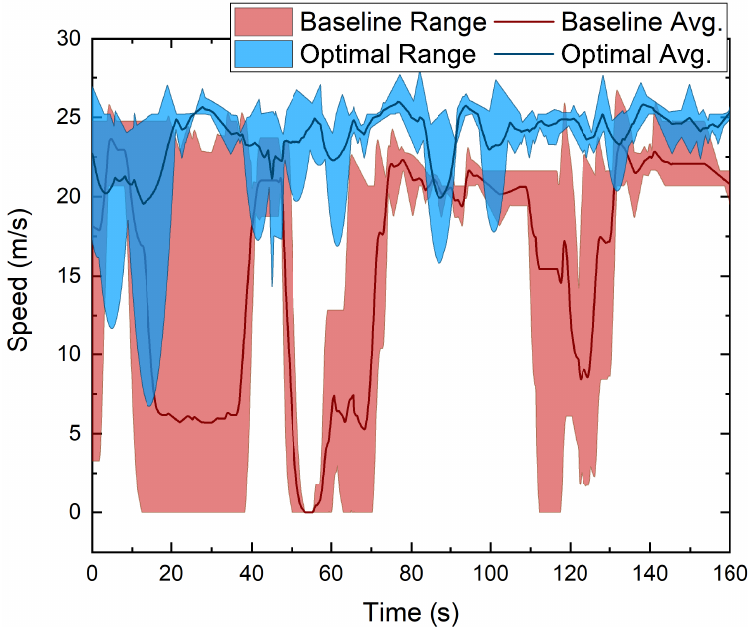}
  } 
  \subfloat[West Loop\label{fig:5-rt4}]{
    \includegraphics[width=0.48\textwidth]{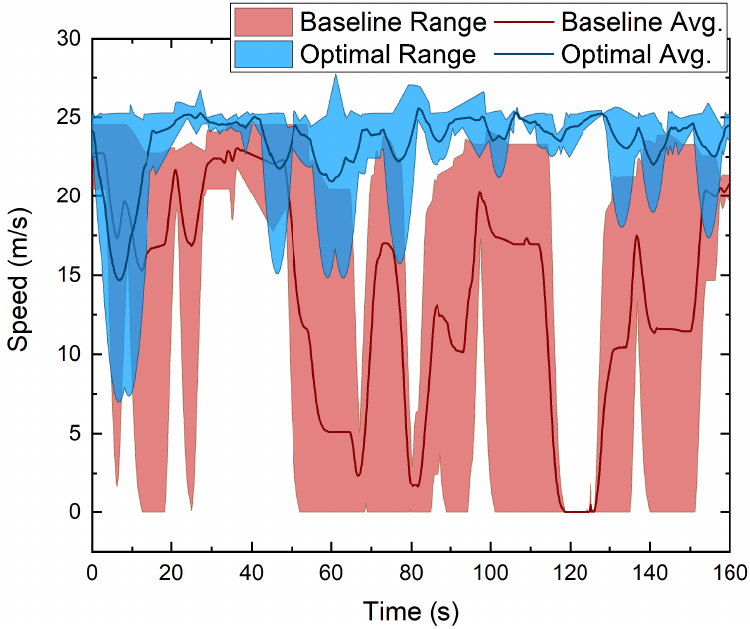}
  } 
  \caption{Instantaneous maximum, minimum, and average speed of each vehicle by route. Each data set is taken from the first $160$s of simulation.}
\end{figure}

The speed profiles of the CAVs, making multiple passes through each loop of the UDSSC map, for the baseline and optimal control scenario are shown in Fig. \ref{fig:5-rt1} - \ref{fig:5-rt4}. The baseline speed profiles show significant stop-and-go driving behavior. The effect of congestion in these cases can induce artificial congestion or \textit{phantom traffic jams} outside the corridor, as can be seen in Fig. \ref{fig:5-rt4} between $60$ and $120$ s.
 
Compared to the baseline speed profiles, 
we note that the 
optimal controller 
has completely eliminated stop-and-go driving, and the optimal quadratic profile speed is realized within the control zones. The congestion observed in the baseline scenario in Fig. \ref{fig:5-rt4} has also been mitigated, rendering an overall smooth traffic flow throughout the network. 
 
Note that the speed 
profiles are kept within the maximum and minimum limit inside the control zone 
as described in Section \ref{sec:simulation-setup}. Therefore, none of the state and control constraints of \eqref{eq:constraints} became active in the unconstrained arc, and our relaxation of \eqref{eq:constraints} holds. However, we see a few cases of constraint violation of the acceleration profile outside the control zone in Fig. \ref{fig:5-rt1} - \ref{fig:5-rt4}, where the Wiedemann car-following model \cite{Wiedemann1974} is applied instead of the optimal control. Note that optimal control is applied only inside the designated control zones.

\subsection{Experimental Validation}

\begin{figure}[htb]
    \centering
    \includegraphics[width=0.8\textwidth]{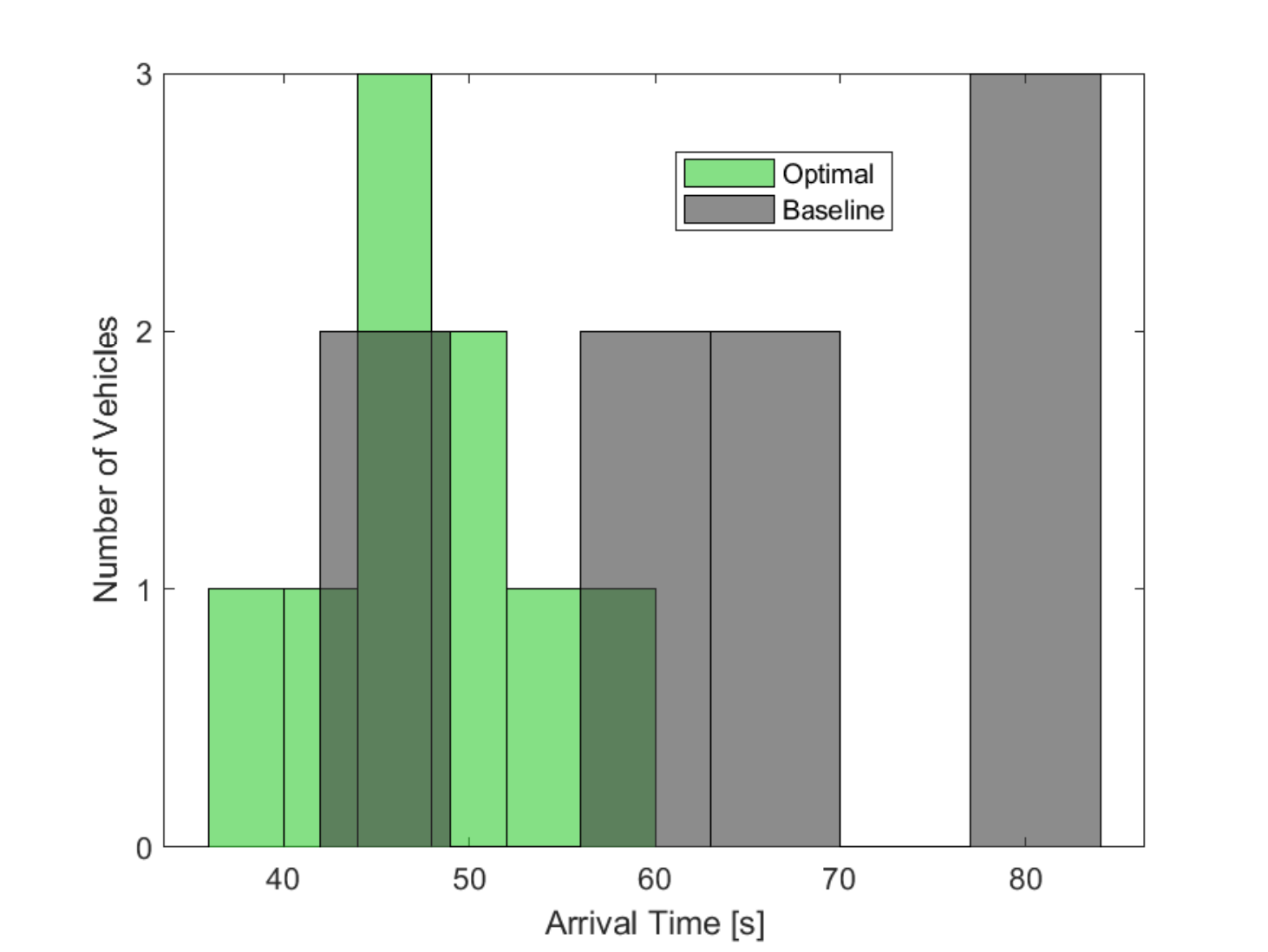}
    \caption{Histogram for the arrival time of each vehicle in Table \ref{tbl:results} with $6$ bins per experiment.}
    \label{fig:time-histogram}
\end{figure}

To validate the effectiveness and efficiency of our optimal control observed in the simulation, we compare the travel time of 9 CAVs  between the baseline and optimal scenarios. The travel time is calculated as the time taken for each vehicle to complete a single loop by sampling the raw VICON data over the $80$ s experiment. In particular, returning to the initial position was defined as the first time the vehicle came within 10 cm of its initial position after a $5$ s initial window. These values are presented in Table \ref{tbl:results} and Fig. \ref{fig:time-histogram} alongside the route each CAV took (as annotated in Fig. \ref{fig:udssc_vissim}); a value of greater than $80$ s corresponds to a vehicle not fully completing its loop during the experiment, which occurred $3$ times in the baseline scenario. Videos of the experiment can be found at the supplemental site, \url{https://sites.google.com/view/ud-ids-lab/tfms}.

\begin{table}[htbp]
\caption{Travel time for each vehicle to complete a single loop.}
\vspace{0.5em}
\centering
\begin{tabular}{ccccccc} \label{tbl:results}
    Vehicle & Baseline time [s]  & Optimal time [s] & Loop & Time saved [s] & \% Decrease \\
    \toprule
        $8$  &  $> 80 $  &   $54.10$  &   North & $> 25.9$ & 32.4 \\
        $18$ &  $> 80 $  &   $48.00$  &   North & $> 32.0$ & 40.0 \\ 
        $12$ &  $66.25$  &   $45.55$  &   East  & $20.70$  & 31.2 \\ 
        $14$ &  $65.95$  &   $49.90$  &   East  & $16.05$  & 24.3 \\ 
        $4$  &  $57.80$  &   $53.25$  &   South & $4.55$   & 7.9 \\
        $13$ &  $60.30$  &   $59.75$  &   South & $0.55$   & 0.9 \\ 
        $2$  &  $46.61$  &   $37.40$  &   West  & $9.21$   & 19.8 \\
        $5$  &  $43.90$  &   $44.20$  &   West  & $-0.30$  & -0.7 \\ 
        $17$ &  $> 80 $  &   $41.50$  &   West  & $> 38.5$ & 48.1
\end{tabular}
\end{table}

From Table \ref{tbl:results}, an average improvement of at least $16.35$ s ($25\%$) over the baseline travel time was observed in the UDSSC. The marginal improvement was in the Southern loop, where the traffic was effectively free-flowing in the baseline scenario. In the loops with conflict zones, i.e., north, east, and west, the impact of the coordinator and optimal control is clear and significant.

\begin{figure}[htb] 
    \centering
    \includegraphics[width=0.9\textwidth]{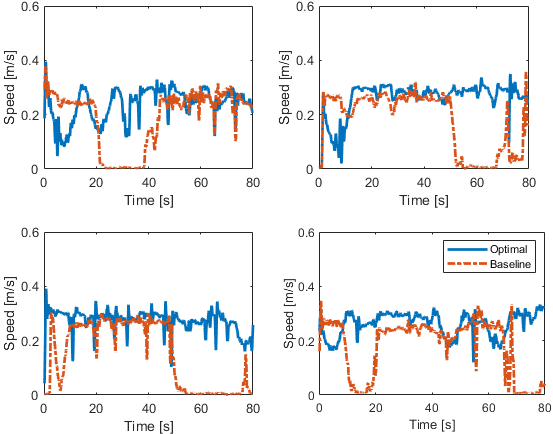}
    \caption{Speed vs time profiles for vehicles (clockwise from top left) 2 (west), 14 (east), 17 (west), 13 (south.).}    \label{fig:vicon-speed}

\end{figure}

The speed profiles of CAVs $2,14,17$, and $13$ are shown in Fig. \ref{fig:vicon-speed}. These profiles were taken by numerically deriving the VICON position data taken at $100$ Hz to get velocity components. Then, any speeds above $0.8$ m/s, well above the maximum speed achievable by the CAVs, was attributed to occlusion during the experiment and thus discarded. Finally, the velocity magnitude was run through a moving average filter with a window of $0.45$ s.

We conclude from the above results that 1) almost the entire reduction in transit time can be attributed to optimal control in the conflict zones, and 2) the optimal control framework almost entirely eliminates stop-and-go driving.

\section{Conclusion} \label{sec: conclusion}

In this paper, we presented an experimental demonstration of a decentralized optimal control framework for CAVs, presented in \cite{Zhao2018}. We used a $1$:$25$ scaled testbed representing an urban environment with robotic CAVs that can replicate real-world traffic scenarios in a controlled environment. We showed that the optimal control framework could contribute a $25\%$ reduction in travel time compared to a baseline scenario consisting of human-driven vehicles without connectivity. We should note that under heavy congested traffic conditions, the control framework might not be as effective as it is under light to medium traffic conditions in improving traffic flow without tuning some control parameters (e.g., minimum time gap, length of the control zone, etc.) \cite{zhao2019}. In recent work \cite{zhao2019a}, the effectiveness of the proposed optimal control framework has been investigated under different traffic conditions to confirm its robustness.

Ongoing research includes the formulation of an upper-level optimization problem, the solution of which yields, for each CAV, the optimal entry time and lane changes required to cross the intersection \cite{Malikopoulos2019b} and explores the associated tradeoffs between throughput and energy consumption of each vehicle. 

An important direction for future research is to consider different penetrations of CAVs, which can significantly alter the efficiency of the entire system. For example, an important question that needs to be addressed is, ``what is the minimum number of CAVs in order to realize potential benefits?" Future work should also consider the robustness of the control framework and its applicability under various traffic conditions. The impact of communication errors and delays on safety and optimality are also areas of future research.

\section*{Acknowledgement(s)}

The authors would like to acknowledge Michael Lashner, Kunzheng Li, Haley Lloyd, Thomas Patterson, and the rest of the UDSSC Senior Design team for their effort in designing, building, and testing the newest generation of CAVs used in this paper. The authors would also like to thank Ioannis Vasileios Chremos for his valuable comments and feedback on the manuscript.

\section*{Funding}
This research was supported in part by ARPAE’s NEXTCAR program under the award number DE- AR0000796 and by the Delaware Energy Institute (DEI). This support is gratefully acknowledged.

\bibliographystyle{tfnlm}
\bibliography{ref,udssc}

\begin{thebibliography}{10}
\providecommand{\url}[1]{\normalfont{#1}}
\providecommand{\urlprefix}{Available from: }

\bibitem{Malikopoulos2016c}
Malikopoulos~AA. A duality framework for stochastic optimal control of complex
  systems. IEEE Transactions on Automatic Control.
  2016;\hspace{0pt}61(10):2756--2765.

\bibitem{Zhao2018}
Zhao~L, Malikopoulos~AA. Decentralized optimal control of connected and
  automated vehicles in a corridor. In: 2018 21st International Conference on
  Intelligent Transportation Systems (ITSC); IEEE; 2018. p. 1252--1257.

\bibitem{Spieser2014}
Spieser~K, Treleaven~K, Zhang~R, et~al. Toward a systematic approach to the
  design and evaluation of automated mobility-on-demand systems: A case study
  in singapore. In: Road vehicle automation. Springer; 2014. p. 229--245.

\bibitem{Fagnant2014}
Fagnant~DJ, Kockelman~KM. The travel and environmental implications of shared
  autonomous vehicles, using agent-based model scenarios. Transportation
  Research Part C: Emerging Technologies. 2014;\hspace{0pt}40:1--13.

\bibitem{Shladover1991}
Shladover~SE, Desoer~CA, Hedrick~JK, et~al. {Automated vehicle control
  developments in the PATH program}. IEEE Transactions on Vehicular Technology.
  1991;\hspace{0pt}40(1):114--130.

\bibitem{Rajamani2000}
Rajamani~R, Tan~HS, Law~BK, et~al. {Demonstration of integrated longitudinal
  and lateral control for the operation of automated vehicles in platoons}.
  IEEE Transactions on Control Systems Technology.
  2000;\hspace{0pt}8(4):695--708.

\bibitem{tsugawa2013}
Tsugawa~S. An overview on an automated truck platoon within the energy its
  project. IFAC Proceedings Volumes. 2013;\hspace{0pt}46(21):41--46.

\bibitem{davila2010sartre}
D{\'a}vila~A, Nombela~M. Sartre: Safe road trains for the environment. In:
  Conference on Personal Rapid Transit PRT@ LHR; Vol.~3; 2010. p. 2--3.

\bibitem{shladover2007path}
Shladover~SE. {PATH} at 20--{History} and major milestones. IEEE Transactions
  on intelligent transportation systems. 2007;\hspace{0pt}8(4):584--592.

\bibitem{Lee2012}
Lee~J, Park~B. {Development and Evaluation of a Cooperative Vehicle
  Intersection Control Algorithm Under the Connected Vehicles Environment}.
  IEEE Transactions on Intelligent Transportation Systems.
  2012;\hspace{0pt}13(1):81--90.

\bibitem{rakha2011eco}
Rakha~H, Kamalanathsharma~RK. Eco-driving at signalized intersections using
  {V2I} communication. In: Intelligent Transportation Systems (ITSC), 2011 14th
  International IEEE Conference on; IEEE; 2011. p. 341--346.

\bibitem{Malikopoulos2017}
Malikopoulos~AA, Cassandras~CG, Zhang~YJ. A decentralized energy-optimal
  control framework for connected automated vehicles at signal-free
  intersections. Automatica. 2018;\hspace{0pt}93:244 -- 256.

\bibitem{mahbub2019ACC}
Mahbub~AMI, Zhao~L, Assanis~D, et~al. {Energy-Optimal Coordination of Connected
  and Automated Vehicles at Multiple Intersections}. In: Proceedings of 2019
  American Control Conference; 2019. p. 2664--2669.

\bibitem{chalaki2019a}
Chalaki~B, Malikopoulos~AA. An optimal coordination framework for connected and
  automated vehicles in two interconnected intersections. In: Proceedings of
  2019 IEEE Conference on Control Technology and Applications, 2019; 2019. p.
  888--893.

\bibitem{Rios-Torres2017}
Rios-Torres~J, Malikopoulos~AA. {Automated and Cooperative Vehicle Merging at
  Highway On-Ramps}. IEEE Transactions on Intelligent Transportation Systems.
  2017;\hspace{0pt}18(4):780--789.

\bibitem{Ntousakis2016aa}
Ntousakis~IA, Nikolos~IK, Papageorgiou~M. Optimal vehicle trajectory planning
  in the context of cooperative merging on highways. Transportation Research
  Part C: Emerging Technologies. 2016;\hspace{0pt}71:464--488.

\bibitem{Malikopoulos2018c}
Malikopoulos~AA, Hong~S, Park~B, et~al. Optimal control for speed harmonization
  of automated vehicles. IEEE Transactions on Intelligent Transportation
  Systems. 2018;\hspace{0pt}.

\bibitem{Athans1969}
Athans~M. {A unified approach to the vehicle-merging problem}. Transportation
  Research. 1969;\hspace{0pt}3(1):123--133.

\bibitem{Kachroo1997}
Kachroo~P, Li~Z. {Vehicle merging control design for an automated highway
  system}. In: Proceedings of Conference on Intelligent Transportation Systems;
  1997. p. 224--229.

\bibitem{Antoniotti1997}
Antoniotti~M, Deshpande~A, Girault~A. {Microsimulation analysis of automated
  vehicles on multiple merge junction highways}. In: IEEE International
  Conference in Systems, Man, and Cybernetics; 1997. p. 839--844.

\bibitem{Ran1999}
Ran~B, Leight~S, Chang~B. {A microscopic simulation model for merging control
  on a dedicated-lane automated highway system}. Transportation Research Part
  C: Emerging Technologies. 1999;\hspace{0pt}7(6):369--388.

\bibitem{Dresner2004}
Dresner~K, Stone~P. {Multiagent traffic management: a reservation-based
  intersection control mechanism}. In: Proceedings of the Third International
  Joint Conference on Autonomous Agents and Multiagents Systems; 2004. p.
  530--537.

\bibitem{Dresner2008}
Dresner~K, Stone~P. A multiagent approach to autonomous intersection
  management. Journal of artificial intelligence research.
  2008;\hspace{0pt}31:591--656.

\bibitem{DeLaFortelle2010}
{de La Fortelle}~A. {Analysis of reservation algorithms for cooperative
  planning at intersections}. In: 13th International IEEE Conference on
  Intelligent Transportation Systems; 2010. p. 445--449.

\bibitem{Huang2012}
Huang~S, Sadek~A, Zhao~Y. {Assessing the Mobility and Environmental Benefits of
  Reservation-Based Intelligent Intersections Using an Integrated Simulator}.
  IEEE Transactions on Intelligent Transportation Systems.
  2012;\hspace{0pt}13(3):1201--1214.

\bibitem{Zohdy2012}
Zohdy~IH, Kamalanathsharma~RK, Rakha~H. {Intersection management for autonomous
  vehicles using iCACC}; 2012. p. 1109--1114.

\bibitem{Yan2009}
Yan~F, Dridi~M, {El Moudni}~A. {Autonomous vehicle sequencing algorithm at
  isolated intersections}. 2009 12th International IEEE Conference on
  Intelligent Transportation Systems. 2009;\hspace{0pt}:1--6.

\bibitem{Li2006}
Li~L, Wang~FY. {Cooperative Driving at Blind Crossings Using Intervehicle
  Communication}. IEEE Transactions in Vehicular Technology.
  2006;\hspace{0pt}55(6):1712,1724.

\bibitem{Zhu2015a}
Zhu~F, Ukkusuri~SV. {A linear programming formulation for autonomous
  intersection control within a dynamic traffic assignment and connected
  vehicle environment}. Transportation Research Part C: Emerging Technologies.
  2015;\hspace{0pt}55.

\bibitem{Wu2014}
Wu~J, Perronnet~F, Abbas-Turki~A. {Cooperative vehicle-actuator system: a
  sequence-based framework of cooperative intersections management}.
  Intelligent Transport Systems, IET. 2014;\hspace{0pt}8(4):352--360.

\bibitem{kim2014}
Kim~KD, Kumar~P. {An MPC-Based Approach to Provable System-Wide Safety and
  Liveness of Autonomous Ground Traffic}. IEEE Transactions on Automatic
  Control. 2014;\hspace{0pt}59(12):3341--3356.

\bibitem{Barth2009}
Barth~M, Boriboonsomsin~K. Energy and emissions impacts of a freeway-based
  dynamic eco-driving system. Transportation Research Part D: Transport and
  Environment. 2009;\hspace{0pt}14(6):400--410.

\bibitem{Berry2010}
Berry~IM. The effects of driving style and vehicle performance on the
  real-world fuel consumption of us light-duty vehicles [dissertation].
  Massachusetts Institute of Technology; 2010.

\bibitem{wu2011}
Wu~C, Zhao~G, Ou~B. A fuel economy optimization system with applications in
  vehicles with human drivers and autonomous vehicles. Transportation Research
  Part D: Transport and Environment. 2011;\hspace{0pt}16(7):515--524.

\bibitem{Malikopoulos2018a}
Zhao~L, Malikopoulos~AA, Rios-Torres~J. Optimal control of connected and
  automated vehicles at roundabouts: An investigation in a mixed-traffic
  environment. In: 15th IFAC Symposium on Control in Transportation Systems;
  2018. p. 73--78.

\bibitem{Rios2018}
Rios-Torres~J, Malikopoulos~AA. Impact of partial penetrations of connected and
  automated vehicles on fuel consumption and traffic flow. IEEE Transactions on
  Intelligent Vehicles. 2018;\hspace{0pt}3(4):453--462.

\bibitem{zhong2017}
Zhong~Z, Joyoung~L, Zhao~L. {Evaluations of Managed Lane Strategies for
  Arterial Deployment of Cooperative Adaptive Cruise Control }. In: TRB Annual
  Meeting; Washington DC, USA; 2017.

\bibitem{Malikopoulos2016a}
Rios-Torres~J, Malikopoulos~AA. {A Survey on Coordination of Connected and
  Automated Vehicles at Intersections and Merging at Highway On-Ramps}. IEEE
  Transactions on Intelligent Transportation Systems.
  2017;\hspace{0pt}18(5):1066--1077.

\bibitem{guanetti2018}
Guanetti~J, Kim~Y, Borrelli~F. Control of connected and automated vehicles:
  State of the art and future challenges. Annual Reviews in Control.
  2018;\hspace{0pt}45:18--40.

\bibitem{wang2018r}
Wang~Y, Li~X, Yao~H. Review of trajectory optimisation for connected automated
  vehicles. IET Intelligent Transport Systems. 2018;\hspace{0pt}13:580--586.

\bibitem{Paull2017}
Paull~L, Tani~J, Ahn~H, et~al. {Duckietown: An open, inexpensive and flexible
  platform for autonomy education and research}. In: Proceedings of the IEEE
  International Conference on Robotics and Automation; 2017. p. 1497--1504.

\bibitem{Hyldmar2019}
Hyaldmar~N, He~Y, Porok~A. A fleet of miniature cars for experiments in
  cooperative driving. In: Proceedings of the IEEE International Conference on
  Robotics and Automation; 2019.

\bibitem{Stager2018}
Stager~A, Bhan~L, Malikopoulos~A, et~al. {A Scaled Smart City for Experimental
  Validation of Connected and Automated Vehicles}. IFAC-PapersOnLine.
  2018;\hspace{0pt}51(9):130--135.

\bibitem{Malikopoulos2019b}
Malikopoulos~AA, Zhao~L. Optimal path planning for connected and automated
  vehicles at urban intersections. In: Proceedings of the 58th IEEE Conference
  on Decision and Control; 2019 (to appear).

\bibitem{Malikopoulos2008b}
Malikopoulos~AA, Assanis~DN, Papalambros~PY. Optimal engine calibration for
  individual driving styles. In: SAE Proceedings, Technical Paper 2008-01-1367;
  2008.

\bibitem{Malikopoulos2019a}
Malikopoulos~AA, Zhao~L. A closed-form analytical solution for optimal
  coordination of connected and automated vehicles. In: Proceedings of 2019
  American Control Conference; 2019. p. 3599--3604.

\bibitem{Wiedemann1974}
Wiedemann~R. Simulation des strassenverkehrsflusses [dissertation].
  Universit{\"a}t Karlsruhe; 1974.

\bibitem{Thrun2007Stanley:Challenge}
Thrun~S, Montemerlo~M, Dahlkamp~H, et~al. {Stanley: The robot that won the
  DARPA Grand Challenge}. Springer Tracts in Advanced Robotics.
  2007;\hspace{0pt}.

\bibitem{Spong2004RobotEdition}
Spong~MW, Hutchinson~S, Vidyasagar~M. {Robot Dynamics and Control Second
  Edition}; 2004.

\bibitem{zhao2019}
Zhao~L, Malikopoulos~AA, Rios-Torres~J. On the traffic impacts of optimally
  controlled connected and automated vehicles. In: Proceedings of 2019 IEEE
  Conference on Control Technology and Applications; 2019. p. 882--887.

\bibitem{zhao2019a}
Zhao~L, Mahbub~AMI, Malikopoulos~A. Optimal vehicle dynamics and powertrain
  control for connected and automated vehicle. In: Proceedings of IEEE
  Conference on Control Technology and Applications; 2019. p. 33--38.

\end{thebibliography}

\end{document}